\documentclass{iau} 
\usepackage{graphicx}
\usepackage{natbib}
\newcommand {\aj} {{\it AJ}}
\newcommand {\apj} {{\it ApJ}}
\newcommand {\apjs} {{\it ApJS}}
\newcommand {\apjl} {{\it ApJL}}
\newcommand {\aap} {{\it A\&A}}

\newcommand {\aapr} {{\it A\&A Review}}
\newcommand {\araa} {{\it ARA\&A}}

\newcommand {\mnras} {{\it MNRAS}}

\newcommand {\nat} {{\it Nature}}

\newcommand {\jqsrt} {{\it J. Quant. Spectrosc. Rad. Transfer}}

\newcommand {\rmp} {{\it Rev. Mod. Phys.}}
\newcommand {\pasj} {{\it Pub. Astron. Soc. Japan}}

\newcommand {\pnas} {{\it Proc. Natl. Acad. Sci.}}

\newcommand {\ssr} {{\it Space Sci. Rev.}}

\def\lapprox{{_<\atop{^\sim}}}

\title[Astrochemistry: overview and challenges] 
{Astrochemistry: overview and challenges}

\author[Ewine F. van Dishoeck]   
{Ewine F. van Dishoeck$^{1,2}$}

\affiliation{$^1$Leiden Observatory, Leiden University \\ P.O. Box 9513,
NL-2300 RA, Leiden, the Netherlands \\ email: {\tt ewine@strw.leidenuniv.nl} 
\\ [\affilskip]
$^2$Max Planck Institute for Extraterrestrial Physics, Garching, Germany }

\pubyear{2017}
\volume{332}  
\jname{Astrochemistry VII - Through the Cosmos from Galaxies to Planets}
\editors{M. Cunningham, T.J. Millar \& Y. Aikawa, eds.}
\begin{document}

\maketitle

\begin{abstract}
  This paper provides a brief overview of the journey of
  molecules through the Cosmos, from local diffuse interstellar clouds
  and PDRs to distant galaxies, and from cold dark clouds to hot
  star-forming cores, protoplanetary disks, planetesimals and
  exoplanets. Recent developments in each area are sketched and the
  importance of connecting astronomy with chemistry and other
  disciplines is emphasized. Fourteen challenges for the field of
  Astrochemistry in the coming decades are formulated.

\keywords{astrochemistry, molecular data, molecular processes,
    ISM: molecules, stars: formation, planets: formation, comets, techniques:
    spectrocopic, galaxies: ISM}
\end{abstract}

\firstsection 

\section{Introduction}

Astrochemistry, also known as molecular astrophysics, is `the study of
the formation, destruction and excitation of molecules in astronomical
environments and their influence on the structure, dynamics and
evolution of astronomical objects' as stated by the pioneer of this
field, Alexander Dalgarno in 2008. This definition covers not only the
chemical aspects of the field, but also recognizes that molecules are
excellent diagnostics of the physical conditions and processes in the
regions where they reside. Moreover, they actively contribute to the
physical state of the gas by being important coolants. From an
astronomical perspective, dense molecular clouds are important as the
nurseries of new generations of stars and planets, some of which may
even harbor life. From the pure chemistry perspective, interstellar space
provides a unique environment in which molecular behavior can be studied under
extreme conditions.  It is this combination that makes Astrochemistry such a
fascinating research field, for both astronomers and chemists alike.

\begin{figure}[b]
\begin{center}
\includegraphics[width=9cm]{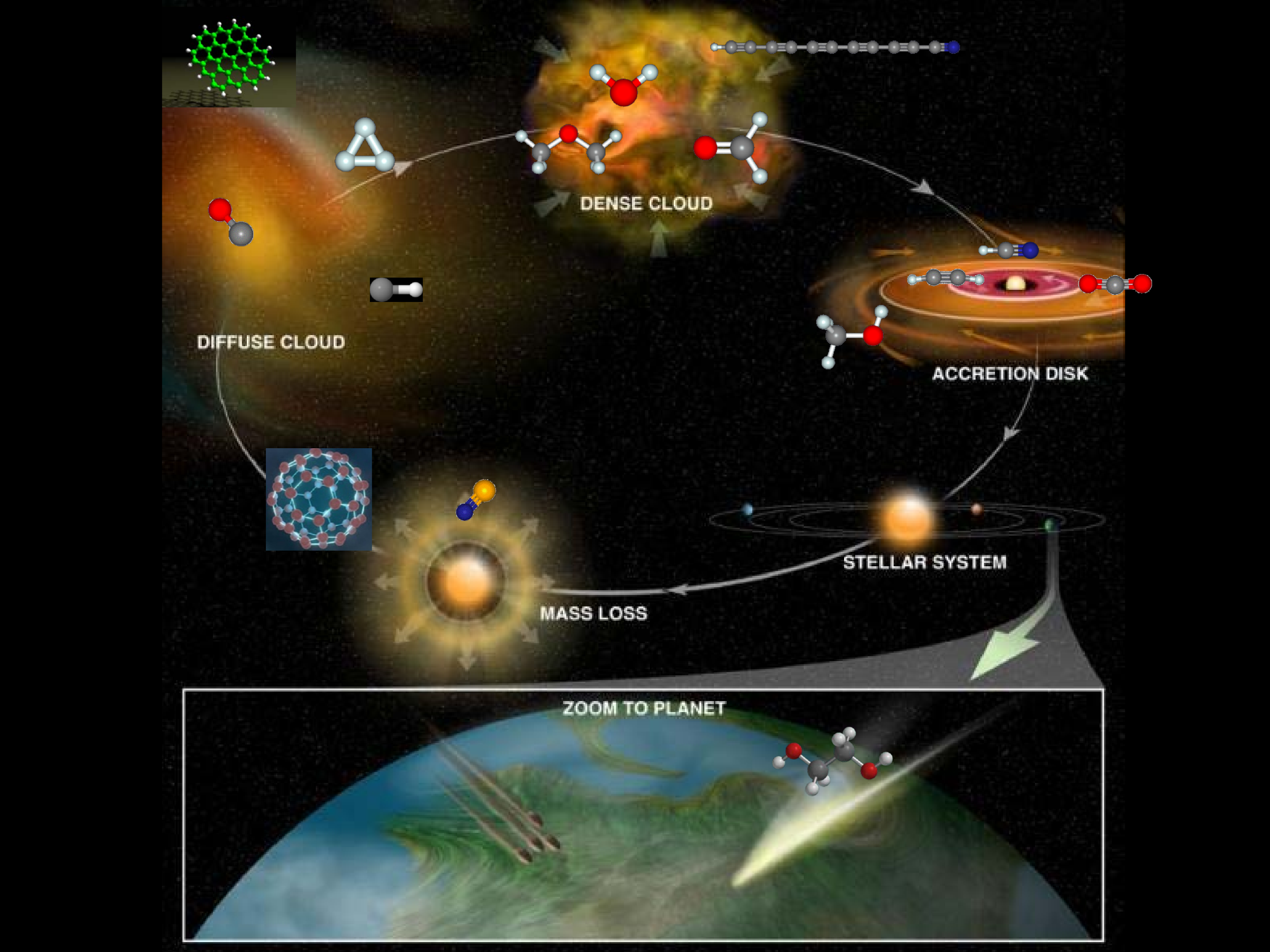} 
\caption{Lifecycle of gas and dust in interstellar space. Some
  characteristic molecules at each of the star- and planet formation
  and stellar death stages are indicated. Image by Bill Saxton
  (NRAO/AUI/NSF) and molecule pictures from {\it the Astrochymist} ({\tt
    www.astrochymist.org}; this website also contains a list of detected
  molecules in space).}
   \label{fig1}
\end{center}
\end{figure}

More than 200 different molecules have been detected in interstellar
space (Fig.~\ref{fig1}).  The main questions in the field of
Astrochemistry therefore include: how, when and where are these
molecules produced and excited?  What do they tell us about
temperatures, densities, gas masses, ionization rates, radiation
fields, and dynamics of the clouds?  How are they cycled through the
various phases of stellar evolution, from birth to death?  How far
does chemical complexity go?  And, most far-reaching, can interstellar
molecules become part of new planetary systems and form the building
blocks for life elsewhere in the Universe?

The interdisciplinary aspects of this field will be emphasized
throughout this overview as a two-way street: Astrochemistry needs
basic data on molecular spectroscopy and chemical processes, but it
also inspires new chemical physics through studies of different
classes of molecules and reactions that are not normally considered on
Earth. Astrochemistry is a `blending of astronomy and chemistry in
which each area enriches the other in a mutually stimulating
interaction' \citep{Dalgarno08}. The field is becoming increasingly
multidisciplinary: besides chemistry, physics and mathematics, which
have traditionally been an integral part of the field, knowledge about
biology, geochemistry and informatics becomes important as well.

In the following, a brief overview of recent developments in the large
range of topics addressed in this Symposium will be given, together
with a challenge for each of these topics. References will be limited
to just a few per topic. Recent overviews of astrochemistry with
extensive reference lists can be found in
\citet{Caselli12aar,Tielens13,vanDishoeck14} and the 2013 Chemical
Reviews special issue, as well as in the individual papers in this
volume.

\section{Observational facilities}

Progress in Astrochemistry is driven by new observational data.
Molecules can be observed through their electronic, vibrational or
rotational transitions at optical/UV, infrared and millimeter
wavelengths, respectively. Astrochemistry has been fortunate that
several of the most powerful new telescopes deployed over the past
decade have been particularly well suited to observe interstellar
molecules: the {\it Spitzer Space Telescope} and the {\it Herschel
  Space Observatory} at mid- and far-infrared wavelengths, a number of
ground-based 8m optical/infrared telescopes equipped with high
resolution spectrometers, and various single-dish and interferometers
at millimeter wavelengths with increasingly sensitive broad-band
detectors, culminating in the Atacama Large Millimeter/submillimeter
Array (ALMA). The {\it Stratospheric Observatory for Infrared
  Astronomy} (SOFIA) is contributing unique data, and the {\it James
  Webb Space Telescope} (JWST), to be launched late 2018, will be the
next big jump in mid-infrared capabilities.

On the longer term horizon the situation looks less favorable for
Astrochemistry, however. ALMA is expected to be still going strong for
decades, and at least one Extremely Large Telescope (ELT) will be
equipped with a high resolution infrared spectrometer. However,
several new flagship space missions in the 2020s--2035s are dedicated
to large area cosmological and galaxy evolution surveys, high energy
astrophysics and to the emerging field of gravitational wave
astrophysics. Exoplanet atmosphere studies will likely be a strong
driver for future satellites, but their modest spectral
resolution instruments have limited applications in other areas of
Astrochemistry.

\smallskip

{\it {\underline {Challenge 1}}: To secure new facilities for
    Astrochemistry in the 2030--2040 timeframe}.

\section{Laboratory astrophysics}
\label{sect:lab}

The availability of accurate atomic and molecular data is another
prerequisite for Astrochemistry.  Note that the term `laboratory
astrophysics' implies experiments as well as quantum chemical and
molecular dynamics calculations. Over the past decade, several new
techniques and new groups have entered the field, ensuring new efforts
to address the many chemical-physics questions and the transfer of (often
unique) expertise to the younger generation. At the same time,
traditional techniques and data gathering remain important as well,
even if some of those experiments are not as `flashy' as using, for
example, He-droplets to study ultracold reactions. The key for a
fruitful interaction is for astronomers to ask the right questions and
have enough chemical physics knowledge of what can, and what cannot,
be done in the laboratory. Chemists, in turn, have to sometimes let go
of perfectionism, since astronomers are often content at the factor of
2 uncertainty level. Also, astronomers just want a number rather than
a deep understanding of the underlying molecular processes. This
`tension' needs to be recognized by both sides, and, importantly, also
by the funding agencies. Equally important, astronomers need to
properly cite the relevant data used in Astrochemistry to support
laboratory scientists.

{\it Spectroscopy:} The most basic information that astronomers need
is spectroscopy from UV to millimeter wavelengths. Techniques range
from classical absorption set-ups to cavity ringdown spectroscopy,
(chirped-pulse) Fourier transform microwave spectroscopy and THz time
domain spectroscopy.  Transition frequencies and strengths of (sub)mm
transitions are summarized in the Jet Propulsion Laboratory catalog
(JPL) \citep{Pickett98} and the Cologne Database for Molecular
Spectroscopy (CDMS) \citep{Endres16} \footnote{{\tt spec.jpl.nasa.gov}
  and {\tt www.astro.uni-koeln.de/cdms/catalog}}.  Databases for
vibrational transitions at infrared wavelengths include the HITRAN
\citep{Hitran09} and the EXOMOL \citep{ExoMol16} line
lists\footnote{{\tt www.cfa.harvard.edu/hitran} and {\tt
    www.exomol.com}}.

Spectroscopy of large samples of PAHs has been mostly carried out with
matrix-isolation experiments, although selected PAHs have also been
measured in the gas phase \citep{Boersma14}. Spectroscopic databases
of solids include those for ices in Leiden, NASA-Ames, Goddard and
Madrid \footnote{{\tt icedb.strw.leidenuniv.nl}}; and the
Heidelberg-Jena-St.\ Petersburg database of optical constants for
silicates \citep{Henning10}. Information on carbonaceous material is
more scattered in the literature.

{\it Rate coefficients:} Chemical models require rates for thousands
of two-body reactions under space conditions.  There are only a
handful types of reactions, however. In the gas-phase, these are
radiative association and associative detachment (formation of bonds);
ion-molecule, neutral-neutral and charge-transfer reactions
(rearrangement of bonds); and dissociative recombination and
photodissociation processes (destruction of bonds).

Gas-phase processes have been summarized by \cite{Smith11} and Sims,
this volume, and many recent developments are described in the 2014
release of the KIDA database \citep{Wakelam15}.
There has been good progress on improving many rate coefficients for
ion-molecule and neutral-neutral reactions over a range of
temperatures. Radiative association, however, is an example of a
process whose rate coefficients continue to have large uncertainties
due to their low values and the difficulties of measuring or computing
them. Photodissociation and photoionization rates of atoms and
molecules exposed to different radiation fields have recently been
updated by \citet{Heays17}.

Rate coefficients for collisional excitation of molecules continue to
be essential for the quantitative non-LTE excitation analysis of the
observed molecular lines. The vast majority of these rates come from
theoretical calculations and the field is fortunate to have a steady
stream of new results from quantum chemists
\citep{Roueff13,Dubernet13,Wiesenfeld16}. With new instrumental
developments on the horizon, it may be time for experiments to catch
up in this area.

{\it Ice chemistry:} The last two decades have seen significant
progress in ice astrochemistry studies, with modern surface science
techniques at ultra-high vacuum conditions now used to quantitatively
study various chemical processes in ices.  Summaries of activities
across the world are given by \citet{Allodi13,Linnartz15}.  Molecules
such as CH$_3$OH, H$_2$O, CO$_2$ have been demonstrated to form at low
temperatures on surfaces through atom addition reactions, validating
reaction schemes postulated 30 years earlier
\citep{Tielens82}. Binding energies are key to many astrochemical
models and more work on astrochemically relevant species and mixtures
is needed \citep{Collings04,Cuppen17}.

Many laboratory experiments across the world have been performed using
some form of `processing' that ices are expected to undergo in space:
heating, UV irradiation, and exposure to energetic particles and/or
electrons \citep{Oberg16}. All of them produce complex molecules, in
some cases even biologically interesting species like amino acids,
sugars, and amides, and often in similar amounts.  The main
differences in outcome may be related to whether or not the strong CO
or N$_2$ bonds can be broken, especially if there is no other source
of atomic C or N in the ice mixture. For example, UV radiation
longward of 1200 \AA \ does not photodissociate CO or N$_2$.
Otherwise the results seem to mainly depend on the amount of energy
deposited in ice, not on which form.  In that respect it is useful to
recall that UV radiation deposits 10$\times$ more energy per molecule
in the ice than cosmic rays for typical cosmic ray fluxes in dark
clouds \citep{Shen04}.  While products are readily identified, the
step-by-step kinetics and mechanisms are still poorly constrained in
most of these experiments, so the results cannot be readily
incorporated into astrochemical models.

Molecular dynamics techniques can also be extended to solid-state
processes, as demonstrated by the detailed study of the
photodesorption of water ice and its isotopologs, as well as CO$_2$
ice formation \citep{Arasa15,Arasa13}. Such models provide much insight
into the processes occuring in and on the ice.

\smallskip

{\it {\underline {Challenge 2}}: To continue to bring chemists,
  physicists and astronomers together to characterize and quantify
  molecular processes that are at the heart of Astrochemistry, to have
  open lines of communication to prioritize needs, and to convince
  funding agencies to continue supporting this interdisciplinary
  research.}

\section{Astrochemical models}
\label{sect:models}

Compilations of reaction rate coefficients together with codes that
solve the coupled differential equations to compute abundances include
the UMIST 2013 database \citep{McElroy13}, and the KIDA database
\citep{Wakelam12} \footnote{{\tt www.udfa.net} and {\tt
    kida.obs.u-bordeaux1.fr}}. The latter website includes the Nahoon
(formally known as Ohio State) gas-grain chemistry code by Herbst and
co-workers.  The UCLCHEM code is now also publicly released
\citep{Holdship17}. These databases and codes contain primarily
two-body reactions because of the low densities in interstellar space.
Three-body reactions become significant at densities above $\sim
10^{13}$ cm$^{-3}$ such as encountered in the inner midplanes of
protoplanetary disks and the atmospheres of stars and exoplanets.
Here the chemistry may approach that in thermodynamic equilibrium
(LTE).

It is important to recognize that astrochemical databases are put
together and updated on a `best effort' basis. Following the good
example of atmospheric chemistry, some databases (UMIST, KIDA) have an
(estimated) uncertainty associated with each rate coefficient which
can then be propagated in the network. Examples of pure gas-phase
networks show that even the abundances of simple molecules like
H$_2$O, SO or CH, have an uncertainty of a factor of $\sim$3 just from
the uncertainties in individual rate coefficients under dark cloud
conditions \citep{Wakelam10b}; for larger molecules, the cumulative
effect of many reactions will be even larger. This illustrates the
level of agreement one can expect between different models, and
between models and observations.

Gas-grain chemistry networks introduce another level of
complexity. The overall efficiency depends on the probability that the
atoms or molecules stick to the grains upon collision, their mobility
on the surface, the probability that molecule formation occurs, and
finally the probability that the molecule is released back into the
gas phase.  The standard treatment of the interplay between gas phase
and grain surface chemistry is through rate equations
\citep{Hasegawa93,Garrod08}. This approach is known to be inadequate
under some conditions, however, especially for models with very small
grains and only a few species per grain.  Many alternative approaches
are being considered such as the modified rate equations, Monte Carlo,
Master equation, and hybrid methods, each with their advantages and
drawbacks.

Chemistry occurs not just at the surface but also deep inside the ice,
since UV photons typically penetrate at least 50 monolayers.  The
atoms and radicals following photodissociation (or some other form of
energetic processing) are created with excess energy and are thus
initially highly mobile, reacting with neighboring species and
overcoming energy barriers. However, they quickly lose this energy on
pico-second timescales and become trapped. Alternatively, these
radicals become mobile and find each other once the ice temperature
increases from $\sim$10 K to 20--40 K.  A general concern is how to
translate laboratory data to parameters that can be used in grain
surface chemistry. Laboratory experiments provide rates as function of
H or UV fluence or some other parameter over timescales of hours;
astronomical applications involve timescales $>10^5$ yr
\citep{Cuppen17}.

Which of these gas-phase or gas-grain reactions dominates the
formation of a molecule depends on the astrophysical situation. For
example, water is primarily formed by ion-molecule reactions in
diffuse clouds; by high temperature neutral-neutral reactions in
shocks; and by grain-surface chemistry in dense cores
\citep{vanDishoeck13}.

\smallskip {\it {\underline {Challenge 3}}: To build realistic gas-grain
  models from microscopic to macroscopic scales, including translation
  of laboratory ice chemistry experiments into parameters that can be
  adopted in models.}

\section{Diffuse and translucent clouds} 
\label{sect:diffuse}

Astrochemistry started nearly a century ago with optical absorption
spectroscopy of atoms and molecules in diffuse clouds along the lines
of sight to bright stars. These clouds, and their somewhat denser
translucent counterparts, have visual extinctions of a few mag and are
in the interesting regime where the transition of most elements from
atomic to molecular form takes place
\citep{vanDishoeck89,Snow06}. Because UV photons and cosmic rays can
penetrate the clouds to ionize atoms and dissociate molecules,
timescales to reach equilibrium are short (few thousand yr) and the
chemistry is dominated by ion-molecule reactions.  Physical parameters
such as density, temperature and UV field are well constrained and
even the H$_2$ column density can be measured directly through UV
absorption lines. Thus, these clouds form the best testbed for {\it
  precision Astrochemistry}, at the factor of 2 or better level. At
the same time, because so many parameters are well determined, they
also serve as diagnostics of the physical structure and processes in
the interstellar medium (ISM).

{\it Herschel}-HIFI data have given a new boost to this field
\citep{Gerin16}.  Pure rotational lines of molecules have been seen in
absorption throughout the Galaxy along the lines of sight toward
distant far-infrared sources. Since most of the population is in the
lowest level(s), the conversion from absorption to column density is
accurate and straight-forward. H$_2$ cannot be observed directly at
far-infrared wavelengths, but HF has proven to be a reliable tracer of
its column density because of its particularly simple chemistry. Small
hydrides like H$_2$O are found to have abundences of $5\times 10^{-8}$
with respect to H$_2$, values that are indeed well explained by
low-temperature ion-molecule chemistry.

One of the main {\it Herschel} surprises has been the discovery of
strong OH$^+$ and H$_2$O$^+$ absorption along most galactic lines of
sight and even in high-redshift galaxies. Also the first noble gas
molecule, $^{36}$ArH$^+$, has been detected. These species point to an
H$_2$-poor phase of the interstellar medium containing molecules
(typically H/H$_2>$10) that had not been recognized before. As such,
the abundances of these molecules form important tests for
hydrodynamical simulations of the large scale ISM. They are also
excellent diagnostics of the cosmic ray ionization rate in the low
density ISM, giving a median value of
$\zeta_{\rm H}=1.8\times 10^{-16}$ s$^{-1}$ that agrees well with
previous results from H$_3^+$ data \citep{Indriolo15}. The origin of
this H$_2$-poor phase is interesting by itself: in contrast with other
molecules, timescales for the formation of H$_2$ itself are very long,
up to $10^7$ yr. Both analytical and hydrodynamical models for the
H/H$_2$ structure under non steady-state conditions 
are now being studied \citep{Bialy17} \footnote{{Non
    steady-state (or time-dependent) chemistry is often (incorrectly)
    denoted as non-equilibrium chemistry by hydrodynamical modelers.}}.

On the other hand, there are lingering discrepancies and mysteries
even for these simple clouds: the high abundance of CH$^+$ (and
SH$^+$) requires introduction of additional parameters to capture the
effects of turbulent heating at low (column) densities
\citep{Godard14}; the discovery of increasingly complex molecules with
abundances similar to those in dark clouds is difficult to explain
\citep{Liszt05}; the need to form grains in the diffuse ISM to solve
the dust lifetime problem implies active grain chemistry
\citep{Krasnokutski14}; and the Diffuse Interstellar Bands remain
unidentified (except plausibly C$_{60}^+$, \citealt{Campbell16}),
pointing to the presence of a reservoir of large carbonaceous
molecules. It may well take another century to solve these questions!

\smallskip

{\it {\underline {Challenge 4}}: To obtain a full inventory of the
  chemical constituents of diffuse and translucent clouds, and
  explain -- at the same time -- their chemical simplicity and
  complexity.}

\section{Photon-Dominated Regions (PDRs)}
\label{sect:pdr}

Photon-Dominated or Photo-Dissociation Regions (PDRs) are clouds
exposed to intense FUV radiation (10$^3$-10$^5$ times the standard
interstellar radiation field, ISRF) controlling their heating and
chemistry \citep{Hollenbach97}. The diffuse and translucent clouds
discussed above are low density, low FUV PDR examples. Since most of
the molecular gas in galaxies is in the form of PDRs, studies of their
structure are highly relevant for understanding spatially unresolved
large scale observations. Locally, the layered structure of PDRs, with
PAH emission and carbonaceous radicals peaking closer to the UV source
than saturated molecules, has been beautifully confirmed by recent
observations of the Orion Bar and Horsehead Nebula
\citep{Guzman14,Cuadrado17}. Such studies are also highly relevant for
analyzing emission from the surfaces of protoplanetary disks, which
are even higher density, higher FUV PDR cases.

X-rays can also affect the chemistry, although the chemical structure
of XDRs shows similarities with that of PDRs since X-rays produce UV
radiation through the interaction of the secondary electrons with
H$_2$. X-rays, like cosmic rays, can however penetrate much deeper
into clouds, heating them over a larger volume \citep{Meijerink07}.
While the individual heating and cooling processes in PDRs and XDRs
have been identified decades ago, it is important to realize that the
resulting temperature structures differ considerably from model to
model \citep{Rollig07}.  This in turn affects the predicted excitation
and emission of molecules in those zones such as the high-$J$ H$_2$
and CO rotational lines \citep{Visser12}. These uncertainties limit
their diagnostic potential and are often not acknowledged by the
community.

Large aromatic molecules are readily observed in PDRs, consisting most
likely of a mix of neutral and ionized PAHs that emit strongly in
mid-infrared bands, although there is still an active discussion on
the fraction of aliphatic moieties in the larger carbonaceous
species. PAHs constitute about 5--10\% of the available carbon, with
perhaps trace amounts of nitrogen \citep{Tielens08}. Fullerenes,
C$_{60}$ and C$_{70}$, have also been identified \citep{Cami10}, and
there is evidence for the conversion of large PAHs to fullerenes when
exposed to intense FUV radiation \citep{Berne12}. Large PAHs are
stable against FUV radiation (except some stripping of H atoms), but
smaller PAHs can be destroyed piecemeal, creating C$_2$ or C$_2$H$_2$
fragments. This, in turn, triggers a top-down chemistry in which small
carbonaceous molecules in PDRs may be produced more efficiently from
destruction of larger species rather than bottom-up starting from
carbon atoms \citep{Zhen14}. JWST will be particularly powerful in
imaging these chemical transition zones in PDRs. Destruction of small
carbonaceous grains in the PDR layers of disks by UV and atomic O has
also been invoked to explain why Earth is so poor in carbon
\citep{Anderson17}.

\smallskip

{\it {\underline {Challenge 5}}: To quantify the importance of top-down
  versus bottom-up chemistry in the production of carbon-bearing
  molecules.}

\section{Evolved stars}
\label{sect:agb}

The chemistry in circumstellar envelopes combines many important
elements of Astrochemistry. From inside to ouside: LTE chemistry in
the stellar atmospheres, non-LTE high temperature shock-induced
chemistry in the inner envelope, the formation of dust grains as the
gas cools, grain surface chemistry, and photodissociation and
ion-molecule chemistry at the outer edge \citep{Decin10}. Because the
wind-accelerated material is moving outward, the time-dependent
chemistry and dynamics need to be coupled. Circumstellar envelopes are
good laboratories to test different types of chemistry, because they
tend to be either carbon- (C/O$>$1) or oxygen-rich (C/O$<$1). The
C/O$>$1 situation, which results in very bright lines of carbon-chain
molecules, occurs under only a few other astrophysical situations,
most notably protoplanetary disks (see \S~\ref{sect:disks}).

The C/O dichotomy is not completely black-and-white, however. {\it
  Herschel} confirmed that H$_2$O is present in a number of
carbon-rich envelopes, as found in earlier SWAS data, but ruled out
the interpretation that the water emission orginates in a cold zone of
sublimating Kuiper-Belt-like icy planetesimals.  More likely, the
presence of H$_2$O under carbon-rich conditions points to a clumpy
envelope with photodissociation breaking up CO and liberating
oxygen. ALMA now allows these various chemical zones in many different
molecules to be imaged, zooming in on the dust formation
zone. Together with future high-resolution infrared spectra, these
data will form critical tests of the so-far poorly constrained models
of dust formation, not only in our own Milky Way but also for sources
in nearby galaxies with lower metallicities such as the LMC and
SMC. Dust destruction in the (reverse) shocks may also be seen. This
will be of fundamental importance for understanding the dust life
cycle in the Universe, including the surprising presence of large
amounts of dust at high redshifts: does most of our dust come from AGB
stars or from supernova remnants?

Almost exactly 30 years prior to this symposium, on Feb.\ 24 1987, SN
1987A exploded in the Large Magellanic Cloud. I remember watching the
supernova shine brightly from a hotel in Puerto Varas where we were
vacationing in March 1987 after an observing run at La Silla. It is
exciting to see ALMA now detecting the cold dust formation directly in
SN 1987A, as well as $^{29}$SiO in the ejecta \citep{Matsuura17}. SiO
was actually first observed by its strong infrared emission a year
after the explosion and chemically modeled at that time by
\cite{Liu94}.

\smallskip

{\it {\underline {Challenge 6}}: To nail down the dust formation and
  destruction processes and their efficiencies in the envelopes of
  evolved low- and high-mass stars, for different metallicities.}

\begin{figure}[t]
\begin{center}
 \includegraphics[width=9cm]{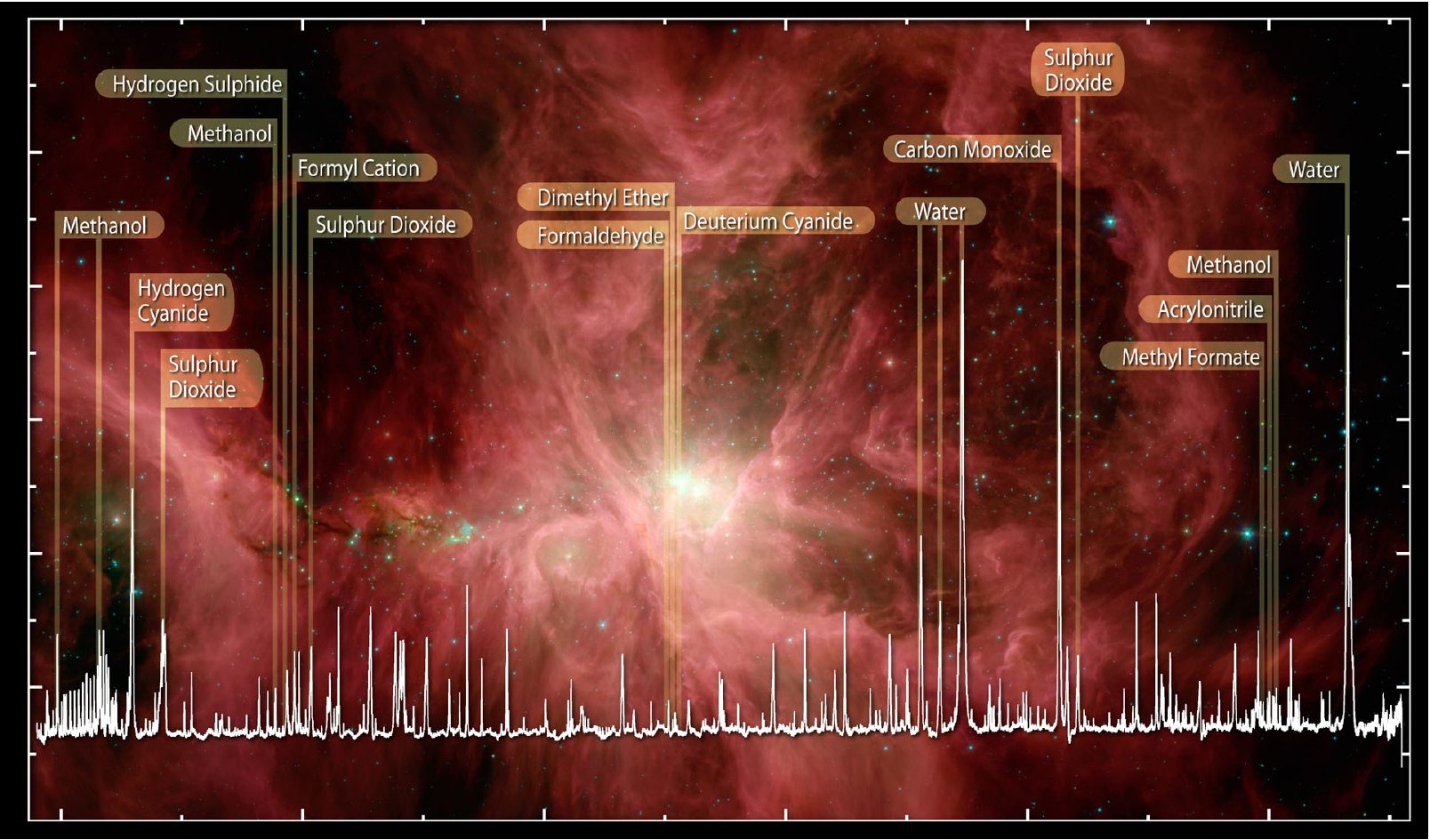} 
 \caption{{\it Herschel}-HIFI spectrum of the Orion KL region around
   550 GHz, showing strong emission from water and organic molecules,
   superposed on a {\it Spitzer} image (NASA/JPL-Caltech/S.T.\ Megeath)
   \citep{Bergin10hexos}.  }
   \label{fig2}
\end{center}
\end{figure}

\section{Extragalactic astrochemistry}
\label{sect:exgal}

Astrochemistry and cosmology go hand-in-hand: the first stars in the
Universe could not have formed without at least some H$_2$ and HD
being present as coolants to allow cloud collapse \citep{Galli13}. This
`dawn of chemistry' \cite[term coined by][]{Dalgarno87} at redshifts
$z$ more than 1000 involves primarily H, D and He and has been well
studied.  It also highlights the importance of accurate rate
coefficients for critical reactions (at the better than 50\% level)
\citep{Kreckel10}. Early Universe chemistry remains outside the
realm of direct observational tests, however.

Chemistry in high-redshift galaxies out to $z\approx 10$ can be observed by
different techniques. Quasar absorption line observations at optical
and millimeter wavelengths probe mostly low-density gas in intervening
galaxies, and reveal H$_2$ and a suite of molecules including OH$^+$,
H$_2$O$^+$ and ArH$^+$ \citep{Muller16} that have been seen in diffuse
galactic clouds (\S \ref{sect:diffuse}).  Thus, {\it Herschel} flew
just in time: without those data, astrochemists would not have been
able to make sense of these extragalactic detections.

Far-infrared and millimeter emission line data probing the denser ISM
show a wide variety of molecules including more complex species
\citep{Gonzalez-Alfonso12,Costagliola15}. Indeed, some of the ALMA
spectra of nearby star-forming galaxies are as rich as those of Orion
were 30 years ago with the poorer sensitivity and spatial resolution
available at that time.

Most studies of extragalactic molecular clouds have been carried out
at kpc scales. ALMA now allows surveys down to scales of $\lapprox$ 20
pc, and exceptionally probing down to 2 pc in nearby galaxies
\citep{Schruba17}. Lensed systems provide an opportunity to image
high-$z$ galaxies on sub-kpc scales \citep{Alma15}. There is a large
range in `metallicities' in these galaxies, with carbon and oxygen
underabundant by factors of 5 to $>$10 compared with solar
abundances. In contrast, most galactic astrochemistry studies have
been done at $<$0.1--1 pc scales in clouds with close to solar
metallicities. Thus, there is a huge gap between galactic and
extragalactic molecular cloud studies that needs to be bridged
\citep{Kennicutt12}; so far, this has not yet been done even for CO,
let alone for other molecules. It is somewhat embarrassing that after
nearly 50 years of millimeter observations, the observed distributions
and abundance ratios in galaxies are such a puzzle.  Cloud-scale
surveys in the Milky Way and nearby galaxies on 1--20 pc scales in key
tracers of different physical structures (e.g., cold gas: N$_2$H$^+$;
warmer gas: SO, CH$_3$OH; dense gas: HCN, HCO$^+$) will therefore
become increasingly important to place the extragalactic data in
context \citep{Pety17,Watanabe17}.

\smallskip

{\it {\underline {Challenge 7}}: To bridge the gap between subpc
  galactic and kpc extragalactic astrochemical studies as functions of
  metallicity out to the highest redshifts, and to use molecular
  observations of calibrated tracers to unveil a new understanding of
  star formation in the early Universe.}

\section{Dense prestellar cores and cold protostellar envelopes}
\label{sect:dark}

Dense cold clouds prior to star formation, with typical temperatures
of 10 K and densities of $10^4 - 10^5$ cm$^{-3}$, exhibit a variety of
chemical characteristics, including long carbon chain molecules, ice
formation and heavy deuterium fractionation \citep{Bergin07}. Most of
these chemical signatures are transferred to the protostellar stage
where they are observed in the outer parts of the collapsing envelope
(Fig.~\ref{fig1}).

{\it Ice formation.} Infrared observations of ice features toward
reddened background stars show that formation of water ice starts in
the translucent cloud phase at $A_V$ of a few mag, when densities are
of order $10^3$ cm$^{-3}$ and dust temperatures $\lapprox 15$~K
\citep{Boogert15}. The formation of water ice is now well
characterized in the laboratory and models, and tested against
observations of cold water gas and intermediate products such as
HO$_2$ and H$_2$O$_2$ \citep[see review by][]{vanDishoeck13}. CH$_4$,
NH$_3$ and some CO$_2$ ice are also made in this early phase, with the
amount of ice rapidly increasing as the cores become more
concentrated. All of these molecules are made on the grain surfaces,
from reactions of atomic O, C and N with atomic hydrogen; they are not
accreted from the gas.

At densities around $10^5$ cm$^{-3}$, the timescales for freeze-out
become shorter than the lifetime of the core, and CO -- the dominant
form of volatile carbon at these high densities -- rapidly depletes
from the gas onto the grains. This `catastrophic' freeze-out of CO
results in a separate `water-poor' or `apolar' ice phase, which has
been observed directly in CO ice profiles. The CO-rich ice can
subsequently react with atomic H to form H$_2$CO and CH$_3$OH, a
process that has been demonstrated to be effective at low temperatures
in the laboratory \citep{Hidaka09,Fuchs09}. Various other routes can
transform CO to CO$_2$ and more complex organic ices.

One of the main observational surprises has been the recent detection
of various complex organic molecules in cold cores
\citep{Bacmann12,Oberg10b1}. The issue is not their production, since
several laboratory experiments have now demonstrated that they can be
formed at low temperatures without the need for heating. The main
puzzle is their return to the gas phase at dust temperatures well
below that for thermal desorption. Non-thermal processes such as
photodesorption, cosmic-ray induced spot heating and reactive
desorption have been proposed, and are being studied theoretically and
in the laboratory, but they remain difficult to quantify and may lead
to desorption of only fragments \citep{Ivlev15,Minissale16,Bertin16}.
Chemical models often still use a `fudge factor' to prevent 
molecules from being fully frozen out: in older models this used to be a
sticking probability of 90--99\% rather than 100\%. In modern models
including detailed grain surface chemistry, a reactive desorption
probability of $\sim$1\% is often introduced to explain observations
\citep{Garrod07}.

\smallskip

{\it {\underline {Challenge 8}}: To identify and quantify the mechanisms
  by which molecules, including the more complex ones, are desorbed
  (intact) from the grain surface in cold clouds.}

\smallskip

{\it Carbon-chain molecules.} A number of cold cores (but not all)
show relatively strong lines of carbon-chain molecules such as
HC$_3$N, HC$_5$N, C$_4$H and CCS, with TMC-1CP still the iconic example
\citep{Kaifu04}. Other cores are more abundant in species like NH$_3$.
One interpretation is that cores rich in carbon chains have only
recently contracted from the diffuse cloud phase and are thus still
rich in atomic carbon, which can be inserted into small molecules to
make longer chains \citep{Suzuki92}. Another explanation is their
environment: if these cores are embedded in filaments or fibers from
which they are still accreting fresh atomic carbon-rich material, they
could build carbon chains through the same process. In the
evolutionary explanation, the statistics on cores with or without
carbon-chain molecules could be used to infer timescales for this
phase.

Cold carbon-chain chemistry should not be confused with the warm
carbon-chain molecule chemistry (WCCC) associated with some low-mass
protostars at a later stage of evolution \citep{Sakai13}. Here one
explanation is sublimation of CH$_4$ ice at moderate dust
temperatures, triggering a gas-phase chemistry that also leads to
carbon-chain molecules.  Alternatively, UV-irradiated outflow cavity
walls can be rich in carbon-bearing molecules such as c-C$_3$H$_2$
(Murillo et al., subm.).

{\it Deuterium fractionation.} Cold clouds have high abundances of
deuterated molecules such as DCO$^+$, DCN and HDCO, with ratios to
their undeuterated counterparts at least three orders of magnitude
higher than the overall [D]/[H] ratio of $\sim 2\times 10^{-5}$.  Even
doubly- and triply-deuterated molecules such as D$_2$CO and ND$_3$
have been detected \citep{Ceccarelli14}.  This huge fractionation can
be explained by two effects. First, the lower zero-point vibrational
energy of deuterated molecules makes their production reactions
exothermic.  In cold cores, most of the fractionation is initiated
by the H$_3^+$ + HD $\to$ H$_2$D$^+$ + H$_2$ reaction, which liberates
230 K.  Correct calculation of the H$_2$D$^+$ abundance requires
explicit treatment of the nuclear spin states (ortho and para) of all
the species involved in the reactions \citep{Sipila10}. At slightly
elevated temperatures, the CH$_2$D$^+$ ion, formed by reaction of
CH$_3^+$ + HD, becomes more effective in controlling the DCN/HCN
ratio.

The second boost in deuterium fractionation occurs in the densest
coldest phase when CO, the main destroyer of both H$_3^+$ and
H$_2$D$^+$, freezes out \citep{Roberts03}.  The increased atomic D/H
ratio in the gas also leads to enhanced deuteration of grain surface
species like H$_2$CO, NH$_3$ and CH$_3$OH
\citep{Tielens83,Furuya16}. Thus, the amount of deuteration in
principle provides information on the formation history. Similarly,
analysis of the size of the freeze-out zones of pre- and protostellar
cores can be used to infer a lifetime of this heavy freeze-out phase
of about $10^5$ yr \citep{Jorgensen05freeze}.

\smallskip

{\it {\underline {Challenge 9}}: To use chemical signatures to
  constrain physical structure, evolutionary stage and the amount of
  time spent in certain cloud phases.}

\section{Shocks, jets and outflows}
\label{sect:shocks}

Shocks with velocities of tens to hundreds of km s$^{-1}$ are common in
the ISM, triggered by a variety of physical phenomena such as
supernova explosions, cloud-cloud collisions as well as jets and winds
from young stars which sweep up surrounding material in outflows. In
the thin shocked layer, gas temperatures increase to a few thousand K,
thus triggering chemical reactions with energy barriers that cannot
proceed in cold gas. Well-known examples are the O + H$_2$ and OH +
H$_2$ reactions leading to the formation of water, each of which has
an energy barrier of about 2000 K. The ubiquitous strong and broad
water line profiles revealed by {\it Herschel}-HIFI in star-forming
regions are a testimony to this high-temperature chemistry
\citep{Kristensen12,SanJose16}.

Shocks can also sputter ice mantles, further enhancing H$_2$O and
other ice mantle molecules like CH$_3$OH into the gas, and at high
velocity even the grain cores themselves \citep{Bachiller99}. Thus,
they can reveal ice mantle material in dark clouds far away from the
driving source \citep{Arce08,Lefloch17}. The release of Si and S leads
to strong emission from molecules like SiO, SO and SO$_2$ associated
with jets. Kinematical signatures in the H$_2$O line profiles
demonstrate the presence of multiple types of shocks that have not yet
been spatially resolved. The observed low H$_2$O abundance, high
OH/H$_2$O ratio and the presence of hydrides like CH$^+$ point to the
importance of UV irradiation of the (pre-)shocked gas and outflow
cavity walls in the chemistry \citep{Karska14Perseus}. All of these
processes are now being imaged at high angular resolution by ALMA,
especially close to the base of the outflows where they originate and
first impact the surrounding envelope.

\smallskip

{\it {\underline {Challenge 10}}: To characterize the chemical and
  physical structure of outflows, especially near the launching point
  of the jets and (disk) winds that drive them.}

\section{Warm protostellar envelopes, hot cores}
\label{sect:hotcore}

The gas and dust close to a protostar are heated by its luminosity,
with temperatures increasing from 10 K in the outer envelope to a few
hundred K in the innermost region. This can result in numerous
chemical changes \citep{Herbst09}: radicals become mobile in and on
icy surfaces and recombine to form even more complex molecules
(so-called first generation complex molecules) and increased UV and
X-rays trigger further chemistry in the ice and gas. When dust
temperatures become high enough for ices to sublimate, molecules do so
presumably in a sequence according to their binding energies. Once
dust temperatures of $\sim$100 K are reached, even the strongly-bound
water and methanol ice sublimate, together with any minor molecules
trapped in them, resulting in particularly rich gas-phase millimeter
spectra (Fig.~\ref{fig2}). This inner 100 K zone is called the `hot
core'.  Here high-temperature gas-phase reactions between sublimated
molecules can result in `second generation' complex organic molecules
\citep{Charnley92}.

Observationally, most chemical surveys of star-forming regions have
still been performed with single-dish millimeter telescopes equipped
with broadband receivers. Over the past decade, pioneering
high-frequency surveys have been carried out with {\it Herschel} of
well-known sources such as Orion \citep{Crockett15}. With ALMA, a new
era has started, since ALMA can image each line {\it and} spatially
resolve the hot core region. Also, ALMA has increased sensitivity to
complex molecules by 1--2 orders of magnitude, and it can study
solar-mass protostars on solar-system scales, not just the most
luminous high-mass sources.  A `sweet' example is the ALMA-PILS survey
of the low-mass protobinary IRAS16293-2422, where an increasing number
of complex molecules is found that had previously only been seen in
SgrB2 and Orion \citep{Jorgensen16}. This includes prebiotic molecules
like glycolaldehyde and ethelyne glycol, and most recently methyl
isocyanate -- a precursor to peptides
\citep{MartinDomenech17,Ligterink17}. In SgrB2, the quest for the
`molecular bricks of life' has now succeeded in the detection of the
first chiral molecule, propylene oxide \citep{McGuire16}, and branched
cyanides such as iso-propylcyanide and N-methyl formamide
\citep{Belloche17}.

Detection of even more complex molecules is becoming hard, however,
even with ALMA, because of the confusion limit. Whether lower
frequency searches with the future SKA or ngVLA provide indeed the
advantages that are claimed still needs to be demonstrated. Also, the
current focus is still on the hunt for individual species in a few
sources; the field should soon move to larger samples with ALMA and
address the deeper underlying questions as to when, where and how
certain classes of complex molecules are produced.

Both single-dish and interferometric surveys show that not all
protostars have line-rich spectra, even though there is evidence for
warm dense dust in their immediate surroundings (e.g., from SED
fitting) \citep{Fayolle15}.  Thus, strictly speaking, not all `hot
cores' have a `hot core chemistry'. Both evolutionary effects and
geometry, such as the presence of a large cold disk on similar scales
as the hot core, as well as beam dilution (for single-dish data) have
been suggested as explanations. Dynamics may also play a role,
especially if infall is fast and molecules spend only a short time in
the hot core region and/or are exposed to excess UV {\it en route}
from cloud to disk \citep{Drozdovskaya16}. Since the time scales of
many of the relevant chemical and dynamical processes in star
formation are similar, there may be subtle effects at play that can
lead to large differences. Coupling chemistry with full (magneto)
hydrodynamical models of star formation is still computationally
challenging and allows investigation of only a limited number of cases
\citep{Hincelin16}.

While complex molecules are traditionally associated only with
protostellar hot cores, they are actually now found at all stages of
star formation: in cold cores (\S~\ref{sect:dark}), shocks
(\S~\ref{sect:shocks}), near PDRs (\S~\ref{sect:pdr}) and in
protoplanetary disks (\S~\ref{sect:disks}). This points to the
increasing importance of forming these molecules under cold conditions
as zero-generation ices without the need for `energetic' or UV
processing, rather than as first- or second generation species. Cold
gas-phase chemistry involving fragments coming off the ice may play a
role in cold clouds and high temperature gas-phase chemistry could
still be important for selected complex species in hot cores
\citep{Taquet16}. The chemical specificity, i.e., the fact that some
complex molecules like HCOOCH$_3$ and CH$_3$OCH$_3$ are more abundant
than others, must hold clues on the relative importance of the various
reaction pathways \citep{Tielens13}.

\smallskip

{\it {\underline {Challenge 11:}} To identify the main formation
  routes of complex molecules in dense clouds, to push detections to
  even higher levels of complexity including prebiotic species like
  amino acids, and to assess how dynamics and geometry during star
  formation can affect their abundances.}

\begin{figure}[t]
\begin{center}
 \includegraphics[width=10cm]{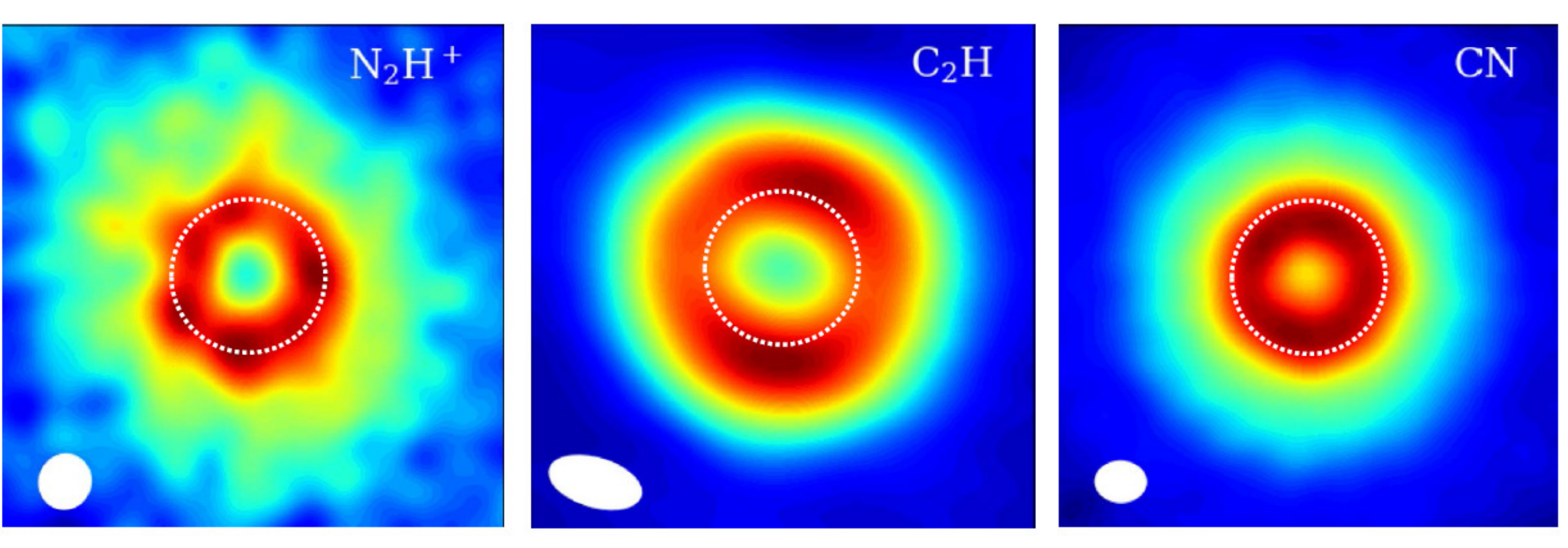} 
 \caption{Molecular rings observed with ALMA in the TW Hya
   protoplanetary disk, showing different locations for each
   species. The white dotted circle indicates the location of the
   N$_2$H$^+$ ring; C$_2$H peaks outside and CN inside this ring
   (Figure by P. Cazzoletti, using data from
   \citealt{Qi13,Bergin16,Teague17}; each image is $6''\times 6''$,
   1$''$=54 AU).  }
   \label{fig3}
\end{center}
\end{figure}

\section{Protoplanetary disks}
\label{sect:disks}

\begin{figure}[t]
\begin{center}
 \includegraphics[width=6cm]{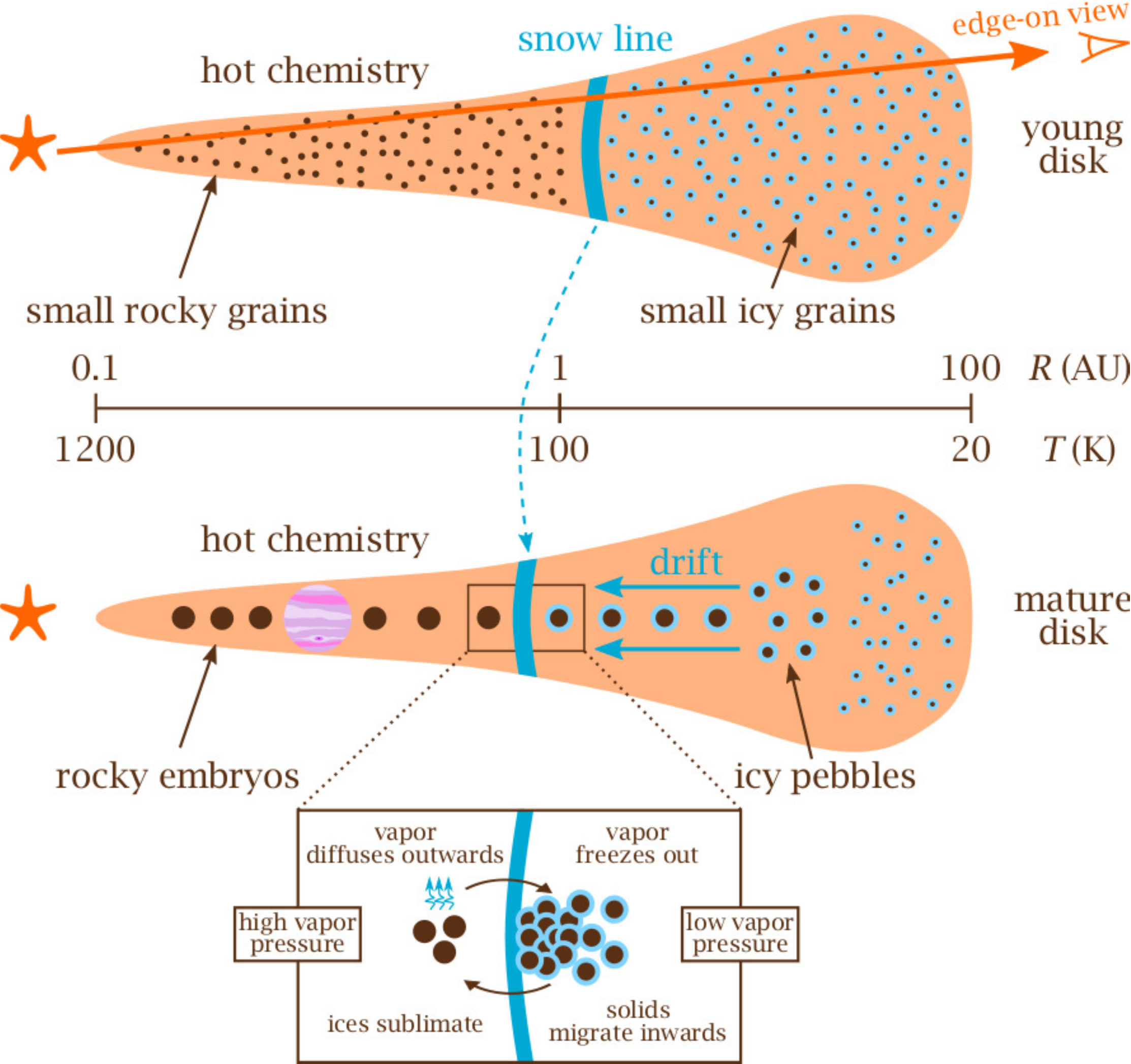} 
 \caption{Sketch of the importance of grain growth, radial drift of
   icy pebbles and snowlines in disk chemistry and in building planets
   and their atmospheres (Figure by R. Visser, after
   \citealt{Sato16}).}
   \label{fig4}
\end{center}
\end{figure}

Protoplanetary disks have become one of the centerpieces of
Astrochemistry, providing the initial chemical conditions for planet
formation. They remain difficult to study both observationally and
theoretically, however. Observationally, signals are very weak since
they are small (typically less than 1$''$ on the sky) and their mass
is only 1\% of that of the collapsing cloud.  Theoretically, they are
a challenge since they cover a huge range of densities and
temperatures in at least two dimensions, from $>$1000 K in the inner
disk and upper layers, to $\leq$10 K in the outer midplane, and from
densities of $>10^{13}$ cm$^{-3}$ in the inner midplane down to $10^5$
cm$^{-3}$ in the upper outer layers. 
Impinging UV radiation fields from the central star can be as high as
$10^5$ times the ISRF in the surface layers at 10 AU, setting up a
PDR-like structure in the vertical direction. Thus, different types of
chemistry are important in different parts of the disks.  Moreover,
gas and dust are largely decoupled (except for the smallest grains):
dust grains grow to pebble size (few cm), settle to the midplane and
drift in radially (Fig.~\ref{fig4}). If they encounter a pressure
bump, dust traps can form where particles can grow to even larger,
planetesimal sizes.  Gas/dust ratios can therefore differ
significantly from 100.

The decreasing temperature in the radial direction sets up a range of
snowlines, i.e., radii where molecules freeze-out onto the grains,
defined as the half-gas, half-ice point. Because of the vertical
temperature gradient, the 2D snow surfaces are actually curved, but
usually only midplane snowline radii are cited. Snowlines are thought
to play a significant role in planet formation, since ice coating of
grains enhances the solid mass and promotes coagulation of grains to
larger particles, effects which are particularly prominent just
outside the snowline. They also control the composition of the icy
planetesimals and gas from which exoplanetary atmospheres are built
(\S 13, 14).

Pre-ALMA observations of disks at millimeter and far-infrared
wavelengths detected mostly simple molecules originating from the
intermediate height warm layers \citep[see][for
reviews]{Bergin07ppv,Henning13}, and confirmed the gas-dust
temperature decoupling predicted by disk models \citep{Bruderer12}.
ALMA is now starting to probe deeper into the disk and spatially
resolve the emission, revealing rings of molecular emission that could
be associated with snowlines. The most obvious example is N$_2$H$^+$,
tracing the CO snowline because its abundance is enhanced when CO is
frozen out (Fig.~\ref{fig3}). However, many other molecular rings are
observed at different locations, with each of them having different
explanations: CO snowlines (N$_2$H$^+$, DCO$^+$), CO photodesorption
(DCO$^+$), chemistry (DCN, DCO$^+$ inner ring), UV (CN), C/O ratios
(C$_2$H, C$_3$H$_3$), gas gaps or cavities (CO isotopologues, CS).
Only two complex molecules, CH$_3$CN and CH$_3$OH, have been detected
so far, and only barely. Even the brightest disks do not show rich
line forests such as found for hot cores. Deep observations of these
and other species in a much larger sample of disks are needed to
understand what the different molecular images are telling us about
the gaseous physical and chemical structure of disks and its relation
to that of the dust. Whether time allocation committees can be
convinced to give large amounts of time to this topic is a challenge
in itself.

The important water snowline is generally out of reach because it
typically occurs at a few AU for T Tauri stars, too small for ALMA to
image. Systems undergoing luminosity outbursts such as V883 Ori
\citep{Cieza16} or young disks with enhanced accretion luminosity
\citep{Harsono15}, for which the water snowlines have moved out to
tens of AU, may offer the best opportunity for direct imaging. More
generally, determining the chemical composition and structure of young
embedded disks may be highly relevant for comparison with comets, if
planetesimal formation indeed starts early (\S~\ref{sect:comets}).

The warm gas in the upper layers of the inner disk ($<$10 AU) emits
strongly at mid-infrared wavelengths. Indeed, a dense forest of lines
due to simple molecules has been detected by {\it Spitzer} and
ground-based infrared telescopes, providing a glimpse of the chemistry
in that region \citep{Pontoppidan14}. Here the high temperatures can
completely `reset' the chemistry. A big advantage of mid-IR
spectroscopy is that key abundant molecules without a dipole moment
such as CO$_2$ and C$_2$H$_2$ can be observed, together with H$_2$O,
OH and HCN, allowing tests of high temperature chemistry. The degree
to which these data can probe the disk midplane chemistry and ice
sublimation, relevant for planet formation models, is still unclear
however, since the mid-infrared dust and line emission is highly
optically thick. JWST will be poised to provide much deeper searches
for CH$_4$, NH$_3$ and minor species, detect isotopologs to constrain
line optical depth, and follow the inner disk chemistry (at
least in the surface layers) from the youngest embedded disks to the
debris disk stage, and across the stellar type range.

\smallskip

{\it {\underline {Challenge 12}}: To make a chemical inventory of
  disks (from inner to outer, surface to midplane, and young to old)
  and relate observed molecular structures to underlying gas and dust
  structures.}

\section{Planet building blocks: comets, planetesimals}
\label{sect:comets}

ALMA and high contrast infrared imaging of disks are providing
striking pictures of the first steps of planet formation: rings, gaps,
cavities, asymmetric structures, and / or spiral arms. Just as for the
molecular rings there are many possible interpretations of these
structures (\S~\ref{sect:disks}), but they do indicate that growth of
dust grains to pebbles and planetesimals is taking place, starting
even in the embedded phase.

How does this affect the chemistry? First hints came from the very
deep {\it Herschel}-HIFI integrations of H$_2$O lines in disks, which
revealed surprisingly weak features \citep{Hogerheijde11,Du17}. The
favorite interpretation is that most water is locked up in large icy
bodies in the midplane and no longer participates in the chemistry at
intermediate disk layers where the gaseous lines can be observed. 
The second clue comes from weak CO isotopologue lines in most disks
\citep{Favre13,Ansdell16}. These lines are reduced in strength not
just due to the well understood processes of freeze-out and isotope
selective photodissociation \citep{Miotello17}, but also by chemistry
transforming CO into CH$_3$OH, CO$_2$ or hydrocarbons
\citep{Yu17}. Alternatively, the rapid planetesimal growth
invoked for H$_2$O also locks up significant amounts of CO and
volatile carbon in various forms.

Thus, the overall picture of the outer disk is that much of the
volatile oxygen and carbon is locked up in ices in large bodies. Since
the dominant H$_2$O and CO$_2$ ices have more oxygen than carbon, the
ice is oxygen rich, with an overall C/O ratio that is lower than the
interstellar or solar abundance \citep{Oberg11}. In contrast, the gas
is carbon rich even though it is overall depleted in carbon and
oxygen. Once C/O$>$1, the same situation occurs as for carbon-rich AGB
stars: the gas has high abundances of small hydrocarbon
molecules. This can indeed explain the strong observed C$_2$H and
c-C$_3$H$_2$ emission in some disks \citep{Bergin16}
(Fig.~\ref{fig3}), although midplane chemistry tends to evolve toward
gaseous C/O significantly less than 1 \citep{Eistrup17}. JWST will be
able to observationally constrain the ice composition in
planet-forming zones, but only for a handful of near edge-on disks and
then only in the intermediate layers.

So how do we determine the disk chemical composition just prior to
planet formation? Much of the chemistry of planet formation is
unfortunately hidden from our view, including the bulk C- and
O-containing species. One option is to observe warmer, younger disks
where less of the material is frozen out and fewer planetesimals have
formed. The forming disks around Class 0 sources, like those for
IRAS16293-2422, may well provide the most detailed information of
chemistry on solar-system scales, even if the material has not yet
settled into a Keplerian structure. That chemistry, in turn, may have
been set already to a large degree in the dense cold core just prior
to and during collapse, arguing for an `inheritance' rather than a
`reset' scenario, at least in the outer disk.

Alternatively, comets and other icy bodies in our own solar system 
provide clues to the chemistry in our natal solar nebula disk, if
they are largely unchanged since their formation 4.5 billion years
ago. Comparisons between cometary and interstellar abundances have
shown some tantalizing similarities \citep{Mumma11}, but there is a
large scatter on both axes. The {\it Rosetta} mission provides an
unprecedented view of the chemical composition of one comet, 67P/C-G
(Altwegg, this volume), which can be compared with that inferred for
protostars such as IRAS16293-2422. However, this is just one comet and
one protostar; ultimately, one would like at least 10 Rosetta-type
missions to comets originating in different parts of our
solar system! More bright comets like Hale-Bopp
that can be probed with ALMA and infrared instruments, including
determining their D/H ratio in water, will provide further insight into
their origin.

\smallskip

{\it {\underline {Challenge 13}}: To determine the (bulk) chemical
  composition and origin of planet-forming material (inheritance or
  reset) and relate that to what is found for icy bodies in our own
  Solar System.}

\section{Exoplanetary atmospheres}

The chemical composition and origin of exoplanet atmospheres, from
Super-Earths to mini-Neptunes and Jovian planets, is clearly the next
frontier for Astrochemistry. So far, only CO and H$_2$O have been
robustly detected, with hints of CH$_4$, NH$_3$ and/or HCN
\citep{Madhu16}. Clouds and hazes often obscure optical and
near-infrared features, especially for the lower mass planets. Even
when features are detected, inferring accurate absolute abundances is
a challenge, although relative abundances are more reliable. For those
planets with clear features, a C/O ratio $< 1$ has generally been
inferred, with so far only one potentially carbon-rich planet.

Exoplanet chemistry models are being developed by various groups, and
vary in sophistication from pure LTE chemistry to the inclusion of
kinetic chemistry, especially photochemistry in the upper parts of the
atmosphere \citep{Venot15}. Titan is a good example of a body in our
own solar system to test nitrogen-rich photochemistry
\footnote{{Photochemistry in the chemical sense, i.e., the fragments
    resulting from photodissociation are produced in excited states
    that react immediately to products before relaxing to the ground
    state. Within Astrochemistry, the term photochemistry should not
    be used for gas phase photodissociation.}}, at the same time also
highlighting its complexity.

One of the ultimate goals is to link the planetary atmosphere
composition with its formation history in the natal protoplanetary
disk. On the one hand, the changing C/O ratio with disk radius could
provide such a probe of the formation location. However, the route
from disk gas+dust to a mature planet is long and involves many steps,
each of them with significant uncertainties \citep{Mordasini16}. For
example, are the heavy elements (i.e., other than H and He) in a giant
planet atmosphere accreted mostly from the gas or delivered by icy
pebbles? How does the migration history affect the outcome?  Moreover,
the disk structure and chemistry is not static but evolving, modifying
the C/O `step function', as does the growth and drift of particles
\citep{Piso15,Eistrup17}. Thomas Henning, at the Kavli Exo-atmospheres
meeting July 2017, lists at least 11 challenges for the field to
overcome, most notably our poor understanding of mass and angular
momentum transport in disks \citep[see also][]{Hartmann17}. It is also
clear that individual molecule abundances will be fully reset in giant
planet atmosphere, preserving only the overall C/O, C/N, O/H etc.\
abundance ratios. And even those could be affected if part of the
atmosphere material is cycled to the planetary core. Thus, as exciting
as this topic is, there are some sobering notes.

\begin{figure}[b]
\begin{center}
 \includegraphics[width=4.5cm]{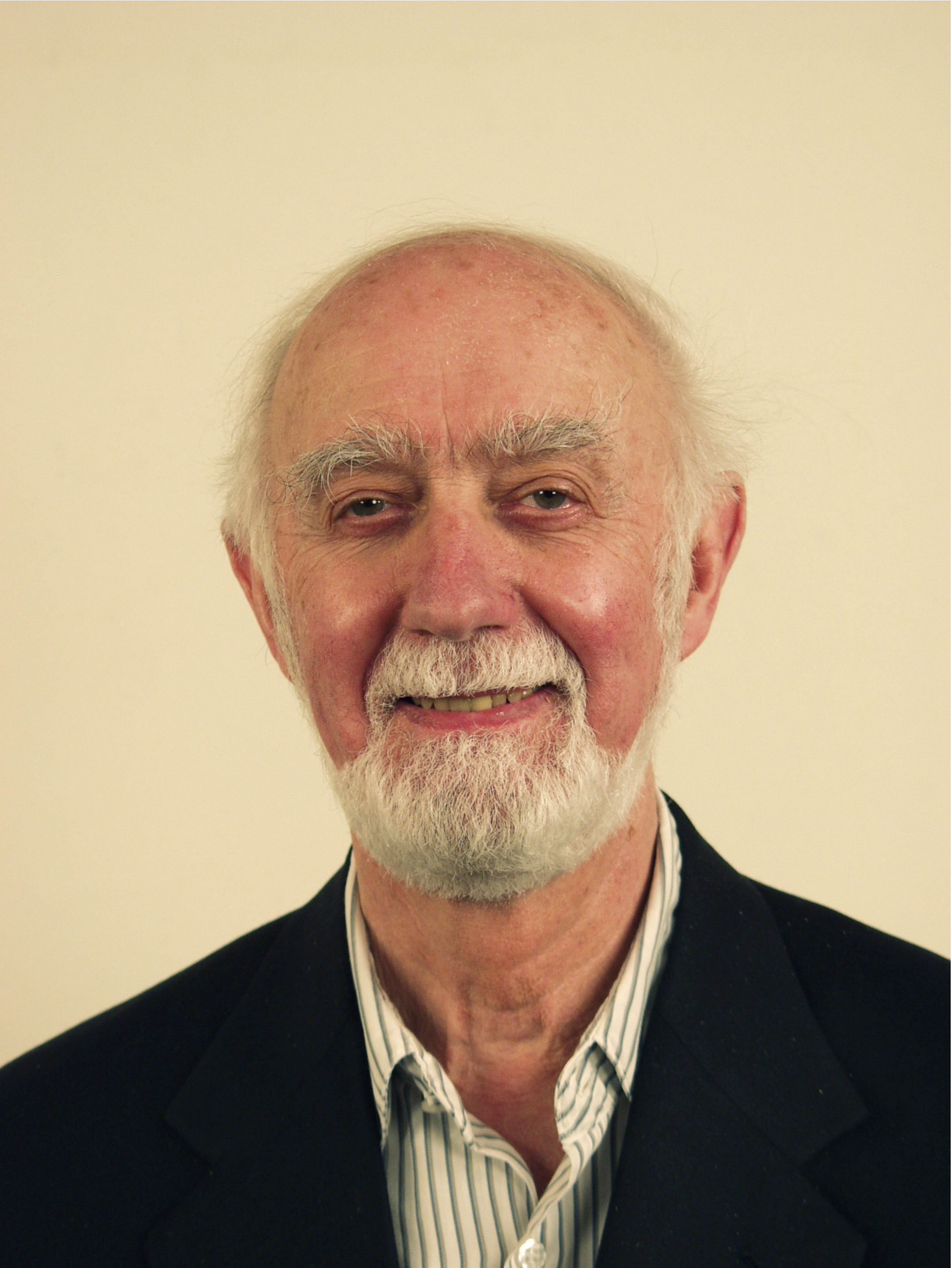} 
 \caption{Alexander Dalgarno in 2003.}
   \label{fig5}
\end{center}
\end{figure}

The atmospheres of rocky terrestrial planets may have an even more
complicated history.  Well outside the water snowline, the planets are
built up largely from planetesimals that are roughly half rock and
half ice.  When these planets move inward, water becomes liquid,
resulting in ‘ocean planets’ or ‘water worlds’.  Inside the snowline,
the planets are usually thought to be very dry. Computing the
atmospheric composition of terrestrial exoplanets is significantly
more complex than that of giant exoplanets and requires consideration
of many additional processes, including even plate tectonics
\citep{Kaltenegger17}.  Because they are very hot initially, they lose
most of their original volatiles through outgassing and atmospheric
erosion. A secondary atmosphere can however be formed through impacts
of (icy) planetesimals in the {\it late} stages of their formation,
including comets such as 67P/C-G which can deliver water and organics
to the young planet. The survival of these organics depends on
planetesimal size and impact speed, and whether the volatile material
can perhaps be shielded by a protective layer on the parent body. If
they do, then there would indeed be a direct link between interstellar
molecules and the building blocks for life on new planets.

Finding signs of life on an alien habitable planet is an ultimate goal
of exoplanet science, and of astronomy as a whole, and is the driver
for the biggest future telescopes and space missions.  Biomarker
molecules have been identified whose detection could indicate the
presence of life on other planets. Key molecules are O$_2$, O$_3$, CH$_4$,
N$_2$O and CH$_3$Cl. However it is also clear that a single molecule
such as O$_3$ is not sufficient but that a combination of features,
including at least one oxidizing and reducing species, is needed
\citep{Kaltenegger17,Catling17}.

In this context, it is interesting that several biomarker molecules,
most notably O$_2$ \citep{Bieler15} and CH$_3$Cl (a molecule produced
on Earth primarily by biological and industrial processes)
\citep{Fayolle17}, have recently been detected in comet 67P/C-G. For
O$_2$, the abundance of 4\% with respect to water makes it the fourth
most abundant ice species. If comets are representative of the icy
planetesimals from which terrestrial planet atmospheres are built, one
should be careful of such `false positives' that can have an abiotic
origin.

\smallskip

{\it {\underline {Challenge 14}}: To determine exoplanetary atmosphere
  compositions and to characterize the chemical changes along the many
  steps to planet formation, necessary to relate exoplanetary
  atmosphere compositions to their birth sites in disks.}

\section{Conclusions}

Looking out over the beautiful Llanquihue lake at the snowline on the
Osorno volcano, the quote attributed to Rabindranath Tagore becomes very
appropriate: {\it `You can't cross the sea merely by standing and
  staring at the water'}. The field of Astrochemistry is vibrant and
poised to address major topics in astronomy. It forms the scientific
basis for one of mankind's biggest questions: `are we alone?'. To
deliver on this promise, however, Astrochemistry has to tackle the
questions and challenges outlined above, do the hard (not
always glorious) work and provide the hard numbers. The field also has
to build the right ships (i.e., new telescopes, new laboratory
experiments, new modeling tools), get the right crew (i.e., excellent
young people with diverse backgrounds), and make the right connections
(i.e., with neighboring fields) to be able to steer the ship in the
right direction and cross the water. This will take time but having
this long term vision ensures that there will be exciting developments
in the coming decades, leading up to Astrochemistry XV!

\medskip

{{\it Acknowledgments.}}  This overview is dedicated to Alexander
Dalgarno (1928--2015), who initiated the IAU working group on
Astrochemistry and organized the first two symposia in this
series. His pioneering work to bring molecular physics into astronomy
inspired many young scientists (including this author) to enter our
field. The broad scope of these meetings reflect the spirit of his
work, having contributed to all aspects of Astrochemistry, and
bringing people from across the world together.

\smallskip

The responses from IAU S332 participants to my question on the future
of Astrochemistry, as well as comments on this draft by E.\ Bergin,
G.\ Blake, E.\ Herbst, T.\ Millar, C.\ Walsh and members of my group,
are much appreciated. The writing of this review was supported by EU
A-ERC grant 291141 CHEMPLAN.







\end{document}